# Inventions on selecting GUI elements
## A TRIZ based analysis

**Umakant Mishra**

Bangalore, India

http://umakantm.blogspot.in

**Contents**

## 1. Introduction

Selecting an object or element is a fundamental operation in any graphic user interface. It is necessary to select an object before doing any operation (such as, dragging, copying, opening, deleting etc.) on that object. The GUI may provide features to select any single object or even multiple objects. The feature of selecting multiple objects can provides tremendous power to the GUI as the user can do complex operations on multiple objects in one go.

However, the process of selection is not as simple as it appears to the user of a GUI. The internal logic of a selection mechanism can be very complex in some situations. Besides there are some fundamental difficulties associated with the selection mechanism, some of which are as below.

⇨ Some objects may be very small or thin that requires very precise location of the cursor for selection. But there is possibility of erratic cursor movements because of uncontrollable movements of the hand and fingers and a minor mistake in cursor movement can result in undesirable results.

- ⇒ If there are more number of selectable elements, the user may have to click that many number of times to select each of them. For example, if there are 30 lines and 40 rectangles in a graphic design, and the user wants to select all the rectangles, then he has to click on each of the 40 rectangles which is not only tedious but also error prone.

- ⇒ The situation becomes complicated if the user wants to select multiple objects of different types.

    For example, if the user selects a line object and a text object, it is difficult to operate on both of them as the line object may have different properties like *thickness* and *arrowhead*, whereas text object may have different properties like *font type* and *style*. The situation worsens if the user wants to select a line object and a network adapter, as they have nothing in common to work with.

- ⇒ While selecting multiple objects, the order of selection may be spatial (such as, according to their appearance on the whiteboard) or temporal (such as, time of their selection or time of creation) or based on any other criteria (such as, name of objects). The method of deciding these orders can be complicated.

- ⇒ In some cases the desired object is not be visible on the whiteboard to select. The user has to explore the object out or scroll the screen to make the object visible before he can select the object. But while exploring the object he may loose his previous screen and previous selections.

- ⇒ Sometimes there are several layers of objects lying one on top of the other occupying the same screen space. In such cases the objects (layers) that remain behind the front object (or layer) are difficult to select. For example, if a line object remains behind a rectangle object, the rectangle object gets selected by the click event when the user tries to select the line object.

- ⇒ It is difficult to select different objects and do various operations on the selected objects by using only one pointer device. Although multiple pointer devices could have been used to increase the efficiency of selection, the user cannot use multiple pointer devices simultaneously by using only one active hand.

Thus there are plenty of difficulties and limitations of a selection mechanism. Let's see how these difficulties have been addressed by different inventions. The following are some inventions on selecting elements in a graphic user interface.

## 2. Inventions on selecting GUI elements

### 2.1 Time-space object containment for graphical user interface (US Patent 5404439)

**Background problem**

In a graphical designing program, the user can create, modify and delete graphical objects on the screen. But there is a general user interface issue of how the user can select objects to operate on. Most user interfaces provide spatial techniques for selecting objects, where the user sweeps across the whiteboard to select all the objects contained inside the swept region. While objects are usually created serially in time by a user, they are typically selected by spatial grouping in a graphical user interface.

**Solution provided by the invention**

US Patent 5404439 (invented by Moral, et al., assigned by Xerox Corporation, issued in April 1995) provided a method where the user can select a group of graphical objects that are meaningfully related to each other. The invention refines the spatial grouping by considering the temporal relationships, and defines algorithms that combine both spatial and temporal grouping to produce an appropriate grouping of objects.

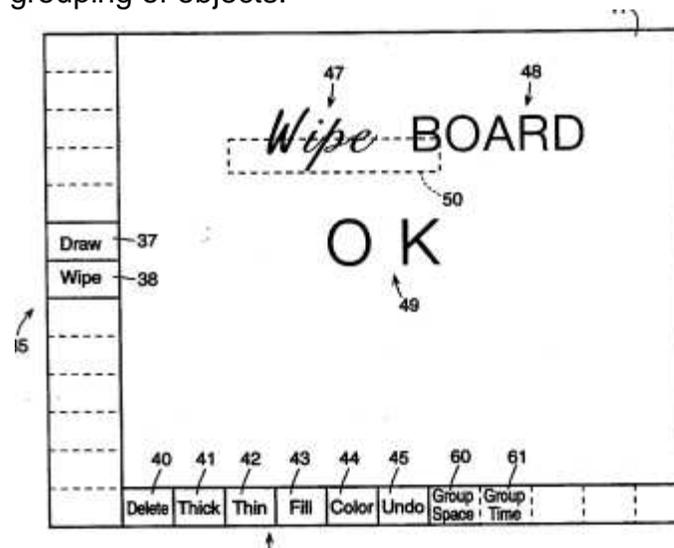

This new method of grouping objects is based on a combination of both spatial and temporal criteria and therefore more likely conforms to the user intentions.

**TRIZ based analysis**

The invention considers including "temporal" criteria for selection of groups rather than only the conventional "spatial" criteria (Principle-17: Another dimension).

It combines both "spatial" and "temporal" criteria to group the selected items (Principle-5: Merging).

## 2.2 Method and apparatus for selecting and displaying items in a notebook graphical user interface (US Patent 5515497)

### Background problem
When the data is presented in notebook pages, the user can scroll up and down to select a desired page in the notebook. But scrolling up and down in a large document takes long time. Besides the operation requires minute observation as otherwise the user might skip the desired page and go on scrolling further. A solution is required to go to a desired page without much scrolling.

### Solution provided by the invention
Patent 5515497 (invented by Itri et al., assigned by International Business Machines Corporation, issued in May 1996) uses a list box to display the items entered on pages of a displayed notebook. According to the invention there will be a relation between the list box items with the notebook page. When the user selects an item in the list box the page having the desired item is immediately displayed on the notebook.

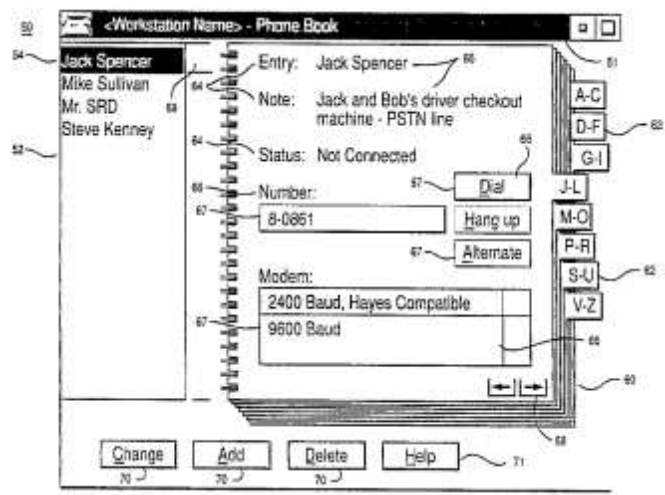

This feature helps the user to select the exact parameter desired without scrolling through a series of displayed pages to find the parameter.

### TRIZ based analysis
The invention uses a list box that displays headlines of all he pages for the user to select a desired page (Principle-24: Intermediary).

The user can select the desired page directly from the list box instead of scrolling through page by page (Principle-21: Skipping).

## 2.3 Refresh and select-all actions in graphical user interface (US Patent 5774120)

### Background problem

The window of a graphical user interface may contain different types of objects, such as, container objects, data objects and device objects. In a graphical user interface, a user often desires to select all the objects or refresh all the objects in a particular window or area. In some currently available graphical user interfaces the select all option is available through a popup menu. But there is no icon/button available for select all or refresh all actions.

### Solution provided by the invention

Patent 5774120 (invented by Goddard et al., assigned by IBM, Jun 1998) proposes a method of refreshing or selecting all objects in a window of a graphical user interface. The invention also provides a graphical means for selecting or refreshing all objects even for windows within windows (generally known as wells).

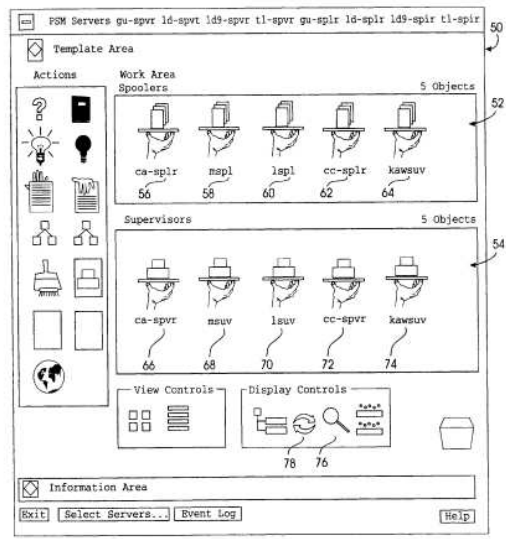

According to the invention, the computer system has a visual button to select all objects or refresh all objects in a window or area. This selection also applies to the windows inside windows.

### TRIZ based analysis

The invention discloses an easy single click method of directly selecting or refreshing all objects in a desired area or window (Principle-21: Skipping, Principle-28: Mechanics substitution).

## 2.4 User interface with simultaneously movable tools and cursor (US Patent 5798752)

### Background problem
Normally the GUI is controlled by a cursor control device (such as a mouse) operated in one hand. There is no use of the other hand although available. This sometimes limits the speed of operation.

### Solution provided by the invention
Buxton et al. Invented a method (Patent 5798752, Assigned by Xerox Corporation, Aug 98) which allows the user to simultaneously work with both the hands. While the non-dominant hand moves the tools, the dominant hand does the finer operations on the worksheet. In a typical implementation the input device may include a trackball (at left hand) for positioning the tools and a mouse (at right hand) for positioning the cursor and initiating actions.

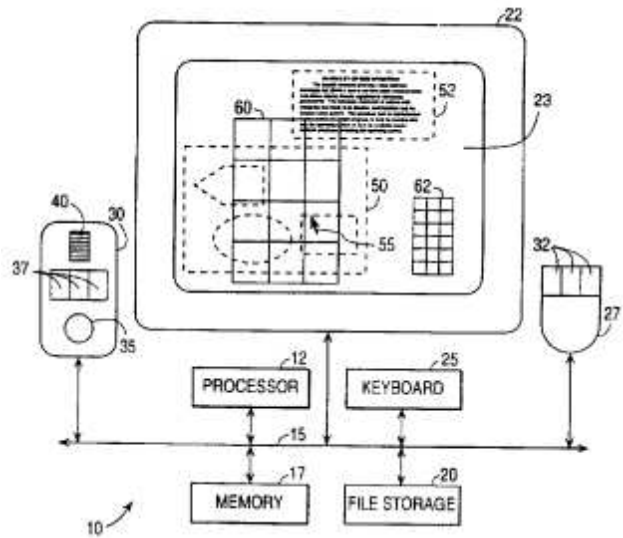

### TRIZ based analysis
The invention advises to make use of the other hand (the less dominant hand which is generally idle) by using another pointing device. (Principle-12: Equipotentiality, Principle-20: Continuity of useful action).

As the efficiency of both the hands are different, the invention suggests to use a trackball (for positioning the tools) with the less dominant hand and a mouse (for positioning cursor and initiating actions) with the more dominant hand. (Principle-4: Asymmetry).

### 2.5 Selection of objects in a graphical user interface (US Patent 6567070)

**Background problem**

In many graphical user interfaces the objects are large enough to be selected by a pointer device like computer mouse. But in some cases the graphical objects may be small or thin to precisely move the mouse onto it.

**Solution provided by the invention**

Patent 6567070 (invented by Light et al., assigned by Intel Corporation, issued in May 2003) discloses a method of selecting graphical objects in a GUI. The method designates an expanded target region in a vicinity of the object in response to determining that the pointer has targeted a location on the object. The method indicates selection of the object when the pointer is in the expanded target region.

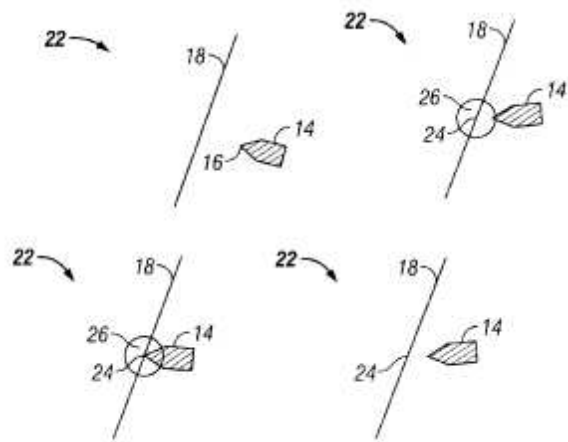

**TRIZ based analysis**

The method uses an expanded target region in the vicinity of the object that helps selection of the object when the pointer is in the expanded region (Principle-16: Partial or excessive action).

## 3. Summary

The difficulties involved in a selection mechanism can be eliminated by various methods. If the object is tiny to select, there can be an expanded target region in the vicinity of the object to simplify selection. If there are multiple objects to be selected, there can be a special button or key-combination to do the job instead of clicking on the objects individually. When the objects are of different types, the selection logic can be intelligent to decide whether to group the clicked object in the selection or not. Similarly, the order in a group of selected objects can be maintained in various ways depending on user requirement and so on. Even one invention proposes to use multiple pointing devises to operate using both the hand to speed up selection process.